\documentclass[12pt, secnumarabic,amssymb, nobibnotes, aps, prd]{revtex4}
\usepackage{epsfig}
\usepackage{amsmath,amssymb}
\usepackage{graphicx}
\usepackage{color}
\usepackage{psfrag}
\usepackage{ulem}

\newcommand{\bneqn}{\begin{equation}}
\newcommand{\edeqn}{\end{equation}}
\newcommand{\ba}{\begin{eqnarray}}
\newcommand{\ea}{\end{eqnarray}}
\newcommand{\bas}{\begin{eqnarray*}}
\newcommand{\eas}{\end{eqnarray*}}

\newcommand{\be}{\begin{enumerate}}
\newcommand{\ee}{\end{enumerate}}
\newcommand{\bitem}{\begin{itemize}}
\newcommand{\eitem}{\end{itemize}}
\newcommand{\barr[1]}{\begin{array}{#1}}
\newcommand{\earr}{\end{array}}
\newcommand{\no}{\nonumber\\}

\usepackage{ulem}

\begin{document}

\title{QCD vacuum and baryon masses  }

\author{Igor A. Mazur}
\email{mazuri@ibs.re.kr}
\affiliation{Center for Extreme Nuclear Matters, Korea University, Seoul 02841,  Korea}
\affiliation{Center for Exotic Nuclear Studies, Institute for Basic Science, Daejeon 34126, Korea}

\author{Youngman Kim }
\email{ykim@ibs.re.kr}
\affiliation{Center for Exotic Nuclear Studies, Institute for Basic Science, Daejeon 34126, Korea}

\author{Masayasu Harada}
\email{harada@hken.phys.nagoya-u.ac.jp}
\affiliation{Department of Physics, Nagoya University, Nagoya 464-8602, Japan}
\affiliation{Kobayashi-Maskawa Institute for the Origin of Particles and the Universe, Nagoya University, Nagoya, 464-8602, Japan}
\affiliation{Advanced Science Research Center, Japan Atomic Energy Agency, Tokai 319-1195, Japan}

\author{Hyun Kyu Lee}
\email{hyunkyu@hanyang.ac.kr}
\affiliation{Department of Physics, Hanyang University, Seoul 133-791, Korea}

\begin{abstract}
To study a possible role of the quantum chromodynamics (QCD) vacuum in nuclear and hadron physics,
we evaluate a physical quantity in a candidate of the QCD vacuum.
In this study we adopt the Copenhagen (spaghetti) picture of the QCD vacuum and calculate the ground-state baryon masses in a constituent quark model.
 We find that the calculated baryon mass does depend on a parameter that characterizes the Copenhagen picture of the QCD vacuum  and satisfies the Gell-Mann-Okubo mass relation for the baryon octet.
 We also observe that the effective constituent quark mass defined in this study contains a contribution attributed to the Copenhagen vacuum, that is the gluon background field. 
 We then estimate the value of the background gluon field as a function of the up (down) constituent quark mass  by using the baryon masses as inputs. 
\end{abstract}
\maketitle

\section{Introduction}
The Quantum Chromodynamics (QCD) vacuum has rich structures which are responsible for chiral symmetry breaking and confinement of QCD: center vortices, instantons, monopoles, quark-antiquark condensate, etc, see ~\cite{Diakonov:2009jq, Greensite:2011zz, Kondo:2014sta, Reinhardt:2018roz, Pasechnik:2021ncb} for a recent review.
A possible role of the QCD vacuum in nuclear and hadron physics has been an important
issue in theoretical physics. For example, the role of the quark-antiquark condensate, which induces spontaneous
 chiral symmetry breaking, has been extensively studied based on
 various effective field theories or models of QCD.

The origin of the nucleon mass is an important issue in theoretical physics, and it is closely related to the QCD vacuum structures.  
The origin of the nucleon mass has been largely attributed to spontaneous chiral symmetry breaking.
For example, in QCD sum rules~\cite{QCDsr}, it was shown that the nucleon mass is proportional to  the quark-antiquark condensate~\cite{Ioffe:1981kw}. This means that the nucleon mass in the chiral limit is solely from spontaneous chiral symmetry breaking, which is  also the case in the linear sigma model.
It is interesting to note that in the parity doublet model~\cite{Detar:1988kn}, the nucleon contains a so-called chiral invariant mass in addition to the contribution from spontaneous chiral symmetry breaking. For example, the nucleon mass in the parity doublet model~\cite{Detar:1988kn} is given by $M_N=\sqrt{(g_1\sigma_0)^2+m_0^2}-g_2\sigma_0$. Here, $g_1$ and $g_2$ are coupling constants whose values are positive. $\sigma_0$ is responsible for spontaneous chiral symmetry breaking and $m_0$ is the chiral invariant nucleon mass.
 Also, the proton (nucleon) mass decomposition based on the energy momentum tensor in QCD has been extensively studied, see~\cite{Lorce:2018uyy, Ji:2021mtz} for a recent review and~\cite{Yang:2018nqn, Liu:2021gco}
for some results from lattice QCD investigations.

In this work, we suggest a way to connect a QCD vacuum property, other than the quark-antiquark condensate, with the baryon mass.
As a minimal setup, we suggest to use the constituent quark model picture in a specific QCD vacuum.
We put the massive constituent quark in a specific QCD vacuum and calculate its energy eigenvalues to obtain the ground-state baryon masses in our approach. For this, we consider a simple minimal coupling between the constituent quarks
and gluons and then solve the Dirac equation for the constituent quarks with
the background gluon field.
As a specific realization of such a  background gluon field, we adopt
the Copenhagen (spaghetti) picture of the QCD vacuum~\cite{Nielsen:1979xu, Ambjorn:1979xi}
which provides a center vortex scenario for confinement\cite{tHooft:1977nqb, Vinciarelli:1978kp, Yoneya:1978dt, Cornwall:1979hz, Mack:1978rq}.
Lattice QCD studies showed that center vortices are also responsible for chiral symmetry breaking in QCD, for example, see~
\cite{deForcrand:1999our, Trewartha:2017ive}. In this connection, The fermion zero modes in the Copenhagen picture were studied in~\cite{Chernodub:2014hla}.
Based on this picture which reveals the advantage of Copenhagen picture, a short description of our suggestion was briefly given in~\cite{Kim:2020ziy}.

In the subsequent section, we introduce briefly the Copenhagen picture of the QCD vacuum and
embed constituent quarks in the Copenhagen picture.
In Sec.~\ref{Bmass}, we solve the Dirac equation of the constituent quarks and evaluate the baryon masses in
the Copenhagen picture of the QCD vacuum.
We then define the effective constituent quark mass in our framework that contains a contribution stemmed from the gluon background field.
We summarize and discuss our results in Sec.~\ref{discussion}.
\section{Constituent quarks in a gluon background  \label{away}}
Understanding the role of rich QCD vacuum structures in nuclear physics is  interesting and important.
A successful and promising way to address this issue is of course the lattice QCD.
However, it is also desirable to have a theoretical tool, based on an effective theory of QCD, to tackle the issue without heavy numerical calculations.

In this section, we study a constituent quark model in the Copenhagen picture.  For some recent discussion about the constituent quark model we refer to
Refs.~\cite{Garcilazo:2001md, Wang:2002ha, Valcarce:2005rr, Inoue:2004jb, Ghalenovi:2019khg, Fernandez:2021zjq}.
Here, we assume that the Copenhagen picture is still valid in the presence of massive constituent quarks.
We begin with a brief summary of  the Copenhagen picture~\cite{Nielsen:1979xu, Ambjorn:1979xi}.
\subsection{A candidate of the QCD vacuum}
There have been enormous amount of studies on the candidates of the true QCD vacuum, center vortices, instantons, monopoles, quark-antiquark condensate, etc,
see ~\cite{Diakonov:2009jq, Greensite:2011zz, Kondo:2014sta, Reinhardt:2018roz, Pasechnik:2021ncb} for a recent review.

As a first step, we begin with a simple classical gluon field configuration, called the Savvidy vacuum.
A constant chromomagnetic field $H$ is a non-trivial classical solution of the SU(2) Yang-Mills equation of motion.
The real part of the one-loop vacuum energy in the constant chromomagnetic field is given by~\cite{Savvidy:1977as, Pagels:1978dd}
\ba
{\rm Re} ~\epsilon=\frac{1}{2}H^2+\frac{11}{48\pi^2}g^2H^2\left(\ln \frac{gH}{\Lambda^2}-\frac{1}{2} \right)\, ,
\ea
 where $g$ is the
gauge coupling constant and $\Lambda$ is a renormalization scale.
This shows that the one-loop vacuum energy has a lower minimum energy at a non-zero value of $H$ than the pertubative QCD vacuum, which implies that quantum fluctuations could generate the homogeneous chromomagnetic field.
Right after this interesting result, a subsequent study~\cite{Nielsen:1978rm} showed
that the Savvidy vacuum is unstable due to the imaginary part in the one-loop vacuum energy,
\ba
{\rm Im} ~\epsilon= -\frac{(gH)^2}{8\pi^2}\, . \nonumber
\ea
The origin of the instability is the tachyonic mode in the lowest Landau level~\cite{Nielsen:1978rm, Huber:1998tr}.
For SU(2) Yang-Mills theory, with a specific choice of the constant chromomagnetic field $G_y^3=Hx$, where
the superscript  denotes the color,
the energy eigenvalue  of the gluon field, $W_\mu= (G_\mu^1+G_\mu^2)/\sqrt{2}$ becomes
\ba
E_n=\sqrt{2gH(n+\frac{1}{2})+k_3^2\pm 2gH}\, .
\ea
Here,  $\pm$ is due to the spin.
Now, one can easily see that the energy eigenvalue can be imaginary for $n=0$ with the minus sign, $E_n=\sqrt{k_3^2- gH}$, when
$k_3^2<gH$.

In Ref. \cite{Consoli:1985sx}, the case of the constant chromomagnetic field  was studied by  using variational techniques
on a class of approximately gauge-invariant gaussian wave functionals, where $H$ is the variational parameter.
It was shown that the minimum of the energy with respect to $H$ is obtained at a non-zero value of $H$.

It is obvious that the constant magnetic field cannot be the true QCD vacuum because it is unstable
and breaks rotational symmetry and gauge invariance. For studies with SU(3) color, we refer to \cite{Flyvbjerg:1980qv, Claudson:1980yz, Chakrabarti:1981zs, Walker:2007uy,Cho:2007ja}.
There have been many studies to remove the instability, for example,
see Refs.~\cite{Nielsen:1979xu, Ambjorn:1979xi, Vercauteren:2007gx, Cho:2012pq, Kondo:2013cka} and references therein.

It was shown in Ref. \cite{Nielsen:1979xu} that the chromomagnetic field has locally
 a domain-like structure and argued that a stable ground state can be obtained by a superposition of
such domains, which restores gauge and rotational invariance. This is
called the Copenhagen vacuum.

In Ref. \cite{Ambjorn:1979xi}, it was argued that the vacuum expectation value of the constant chromomagnetic field is zero everywhere
 in the vacuum which is the quantum liquid state~\cite{Nielsen:1979xu}.
Also, an interesting result was obtained that the coupling constant
that minimizes the vacuum energy of the Copenhagen vacuum is unexpectedly small $\alpha_s=g^2/4\pi=0.37$~\cite{Ambjorn:1979xi}.

\subsection{Constituent quarks in the Copenhagen picture\label{CCp}}
 The starting Lagrangian  for our low-energy effective model
is the same with that of QCD except that the quark mass is not negligible compared to $\Lambda_{\rm QCD}$.
\ba
{\cal L}= \bar\psi i\not\!\!{D}\psi -m\bar\psi\psi
-\frac{1}{2} F_{\mu\nu}F^{\mu\nu}+ \ldots\, ,\label{Lag-1}
\ea
where $D_\mu=\partial_\mu+ig G_\mu$.
Here, $m$ represents the constituent quark mass whose origin is in general believed to be spontaneous chiral symmetry breaking.
 The value of $g$ in our case is different from that in QCD since we are dealing with
the constituent quark. For example, in the case of the chiral quark model~\cite{Manohar:1983md}, the value was
 determined to be $\alpha_s=g^2/4\pi=0.28$. We could not determine the value of it in the present study as explained in the
 next section.
In the constituent quark model, the dominant contribution to the baryon mass is just the sum of the constituent quark masses.
The gluon field is normalized as
\ba
&&G_\mu=G_\mu^a T^a, \no
&& tr(T^aT^b)=\frac{1}{2}\delta^{ab},~~~T^a=\frac{1}{2} \lambda^a\, ,
\ea
where $\lambda_a$ denotes the Gell-Mann matrix.
$F_{\mu\nu}$ is the usual field-strength tensor of the gluon field
\ba
F_{\mu\nu}\equiv \partial_\mu G_\nu -\partial_\nu G_\mu +ig [G_\mu, G_\nu]\, . \nonumber
\ea
We now include  the Copenhagen picture of the QCD vacuum in the following way. In the present work we consider quarks with three colors, for instance,  red, green and blue.  Since our primary goal  is to find out how the vacuum effect is imprinted on the baryon masses, we generalize the simple constant chromomagnetic field used in~Ref.~\cite{Nielsen:1978rm}.

To incorporate the Copenhagen picture in the Dirac equation,
we first define three mutually orthogonal fixed unit vectors in a coordinate space
\begin{align}
\hat{u}_{\parallel} &= (\sin\theta\cos\phi, \sin\theta\sin\phi, \cos\theta) \ ,
\label{u para}
\\
\hat{u}_\perp^{(1)} &= (-\cos\theta\cos\phi, -\cos\theta\sin\phi, \sin\theta)\ ,\\
\hat{u}_\perp^{(2)} & = (\sin\phi, -\cos\phi, 0) \ .
\end{align}
We choose $\hat{u}_\perp^{(2)}$ as the spatial direction of the background gluon. Then,
$\hat{u}_{\parallel}$ denotes the direction of the constant chromomagnetic field as $ \vec{H} = H \hat{u}_{\parallel} = \vec\nabla \times {\vec G}$.
In general, $\vec{H}$ can be defined as $ \vec{H}^a = \vec{\nabla} \times \vec{G}^a - g \epsilon^{abc} \vec{G}^b \times \vec{G}^c$. In the present assumption, however, $\epsilon^{abc} \vec{G}^b \times \vec{G}^c$ is zero since we will turn on the background gluon field
only with $a=3$.

Now, we choose the background gluon field as
${\vec G}_3=Hx_\perp^{(1)} \hat{u}_\perp^{(2)} \lambda_3/2$; all the other gluon fields are purely quantum fields.
Here, $x_\perp^{(1)}={\hat u}_{\perp}^{(1)}\cdot{\vec x}$, and $\lambda_3$ is the third element of the Gell-Mann matrix, $\lambda_3=diag (1,-1,0)$. Then, it is easy to see that the quarks with red and green colors couple to the background gluon field with opposite sign ($gH$ or $-gH$) due to $\lambda_3$, while the blue quark does not couple to the background gluon field.
The eigenvectors of $\lambda_3$ are given by
\begin{equation}
R= \begin{pmatrix} 1 \\ 0 \\ 0 \end{pmatrix},\, G= \begin{pmatrix} 0 \\ 1 \\ 0 \end{pmatrix},\, B= \begin{pmatrix} 0 \\ 0 \\ 1 \end{pmatrix} .
\end{equation}

\section{Baryon mass \label{Bmass}}
In a usual (chiral) constituent quark model, with a confining potential,  we solve the Schr\"odinger equation or Dirac
equation or Faddeev equation to describe the properties of the hadrons. It is widely recognized that the constituent
quark mass contributes dominantly to the ground-state baryon mass.
 Since the
main goal of this work is to show how the QCD vacuum represented by the background gluon fields appears in
physical quantities such as the baryon mass, we assume that the ground-state baryon mass is approximately the sum of three constituent quark masses (or the ground state energies of three quarks). As stated in the previous sections, the Copenhagen picture provides a center vortex scenario for confinement. Therefore, we assume that the constituent quarks are confined in the Copenhagen vacuum.  
We then solve the equation of motion for the quarks in the presence of the constant chromomagnetic field,
considering the fact that the vacuum expectation value of the constant chromomagnetic field is zero everywhere in the Copenhagen picture of the QCD vacuum~\cite{Nielsen:1979xu, Ambjorn:1979xi}.

Now, we solve the Dirac equation for the constituent quark with the constant chromomagnetic field, taking into account the fact that
 at each point one has rotational invariance.
To fulfill this, we need to solve the Dirac equation with  the constant chromomagnetic field $\vec H= H \hat{u}_{\parallel}$,
where  $\hat{u}_{\parallel}$ is defined in Eq.~(\ref{u para}),
and take the average over $\theta$ and $\phi$.

We remark here that there have been an enormous number of works that solves the Dirac equations in a constant field without considering the angle dependence, see Ref.~\cite{Miransky:2015ava} for a recent review. We sketch  briefly how to solve the Dirac equation in our case with the explicit $\theta$ and $\phi$ dependence. After using
\begin{equation}
\psi(x)=(i\not\!{\partial}-g\not\!{G}+m)\Phi(x)\, ,\nonumber
\end{equation}
 we obtain
 \begin{equation}
 \left[(\partial+igG_\mu)^2 +m^2+\frac{1}{2} g \sigma^{\mu\nu}F_{\mu\nu}  \right]\Phi(x)=0\, ,    
 \end{equation}
 where $m$ is the constituent quark mass.
It is convenient to take
$\Phi(x)= c e^{-i k_\parallel\cdot x_\parallel-i k_\perp^{(2)} x_\perp^{(2)}} H_n(x_\perp^{(1)})$,
 where $H_n$ ($n=0,1,2,\ldots$) are the Hermite polynomials for quarks with red and green colors.  Then,
 the kernel for the quark with red color becomes
\begin{equation} 	
\begin{pmatrix}
gH(n+\frac{1}{2})+m^2+gH\cos\theta&-gH\sin\theta e^{-i\phi}\\
gH\sin\theta e^{i\phi}&gH(n+\frac{1}{2})+m^2-gH\cos\theta
\end{pmatrix}.\label{ematrix}
\end{equation}
Since we are interested in the ground state  baryon masses, we have taken $k_\parallel=0$.
The kernels for the quark with green color is obtained by changing $gH$ to $-gH$ in the above expression, and that for the quark with blue color is given with $gH=0$.
We can obtain the Landau level of the constituent quarks by diagonalizing the matrix in Eq. (\ref{ematrix}).
Interestingly, we find that the average over the angles, $\theta$ and $\phi$, are trivial  since  the resulting Landau level of the quark is independent of the angles,
\begin{equation}
E_n=\sqrt{2{\tilde g} H(n+\gamma-1)+m^2 },\quad n=0,1,2,...\, ,\label{qLL}
\end{equation}
where $\gamma=1,2$ is for the spin up and down, respectively, for the quark with red color, and vice versa for the quark with green color. For the quark with blue color, there is no effect from the background field, so that the minimum energy is just $m$.
Here, ${\tilde g} =g/2$ and the factor $1/2$ comes from $G_\mu=G_\mu^a\lambda_a/2$.

To study the effects of the constant chromomagnetic field on the baryon masses,
we assume that the dominant contribution to the baryon mass is the sum of the ground state energy of the constituent quarks and calculate the
ground-state baryon masses.
As an illustrative example, we consider the proton mass in some detail.
The wave function of the proton in the spin-flavor space  is given by
\begin{eqnarray}
&&|p\uparrow\rangle = -\frac{1}{\sqrt{18}}[~uud (\uparrow\downarrow\uparrow+ \downarrow\uparrow\uparrow-2\uparrow\uparrow\downarrow)
+udu(\uparrow\uparrow\downarrow+ \downarrow\uparrow\uparrow-2\uparrow\downarrow\uparrow )\nonumber \\
&&~~~~~~~~~~+ duu (\uparrow\downarrow\uparrow+ \uparrow\uparrow\downarrow-2\downarrow\uparrow\uparrow)~]\, ,
\label{mp}
\end{eqnarray}
where
\begin{align}
uud (\uparrow\downarrow\uparrow+ \downarrow\uparrow\uparrow-2\uparrow\uparrow\downarrow)
= u(\uparrow)u(\downarrow)d(\uparrow)+ u(\downarrow)u(\uparrow)d(\uparrow)-2u(\uparrow)u(\uparrow)d(\downarrow) \ ,
\end{align}
and so on.
The color wave function, which is common to all baryons, is given by
\begin{eqnarray}
\phi_{color}^B = \sqrt{\frac{1}{6}} \left( RGB-RBG+BRG-BGR+GBR-GRB\right)\, .\nonumber
\end{eqnarray}
The contribution from the first (second, etc) term in Eq. (\ref{mp}) with the first term in the color wave function ($RGB$)
is  $3m$ ($m+2\sqrt{m^2+2{\tilde g}H}, \rm{etc}$). Here, $m$ denotes the  up and down constituent quark masses, $m_u=m_d=m\neq m_s$.
After considering all the possible combinations and averaging over seventy-two terms, we obtain $m_{p\uparrow}=2m+\sqrt{m^2+2{\tilde g}H}$.
We confirm $m_{p\uparrow}=m_{p\downarrow}$, as it should be, and  $m_p=m_n=2m+\sqrt{m^2+2{\tilde g}H}$.

In Table \ref{bm}, we summarize our results of the ground-state baryon masses.
Since we calculate the ground-state baryon masses in the constituent quark model, we expect that the Gell-Mann-Okubo mass relation for the baryon octet~\cite{GOr}, $m_\Sigma+3m_\Lambda=2(m_N+m_\Xi)$, will be satisfied.
As it can be seen from Table \ref{bm}, the Gell-Mann-Okubo mass relation for the baryon octet is satisfied.


As manifest in Eq.~(\ref{qLL}), the spin projection along the direction of
the constant chromomagnetic field is an important factor for the eigenvalue of the
constituent quark, especially at the lowest Landau level (LLL).
When $n=0$ and $\gamma=1$, the energy of the corresponding quark does not depend on $H$.
Also, excited states depend on the QCD vacuum more strongly than the LLL, which implies
that the QCD vacuum effect is more influential on the excited baryons than the
ground-state ones.
This observation may be attributed to a specific feature of the
vacuum we chose that provides a constant field to the system.

It is interesting to note here that if we define
\begin{equation}
m\to\tilde{m}=\frac23m+\frac13\sqrt{m^2+2\tilde{g}H},\quad m_s\to\tilde{m}_s=\frac23m_s+\frac13\sqrt{m_s^2+2\tilde{g}H}\, ,
\label{effective mass}
\end{equation}
the baryon masses in Table 1 are equal to the sum of the effective constituent quark masses ($\tilde m$, $\tilde m_s$):
\begin{eqnarray}
	M_N         = 3\tilde{m},\,
    M_\Lambda   = 2\tilde{m}+\tilde{m}_s ,\,
    M_\Sigma    = 2\tilde{m}+\tilde{m}_s ,\,
    M_\Xi       = 2\tilde{m}_s+\tilde{m}\, .	\label{effective mass2}
\end{eqnarray}
This  implies that the effective constituent quark mass $\tilde m$ has a contribution from the gluon background field.
The contribution with $\tilde gH$  is the Landau level energy of the constituent quark in a chiral symmetry broken phase, where
the Landau level is generated by the gluon background field.
We here remark that we have also performed the same analysis for the ground-state meson masses and arrived at the same conclusion.

\begin{table}[!t]
\caption{Calculated baryon masses. $m$ is the mass of $u$- and $d$-quarks and $m_s$ is for the $s$-quark.}
\begin{tabular}{c|c|c}
\hline\hline
Baryon     &Quark contents& Mass                \\
\hline
$p$        & $uud$ & $2m+\sqrt{m^2+2\tilde gH}$ \\
$n$        & $udd$ & $2m+\sqrt{m^2+2\tilde gH}$ \\
$\Sigma^+$ & $uus$ & $\frac43m+\frac23m_s+\frac23\sqrt{m^2+2\tilde gH}+\frac13\sqrt{m_s^2+2\tilde gH}$\\
$\Sigma^0$ & $uds$ & $\frac43m+\frac23m_s+\frac23\sqrt{m^2+2\tilde gH}+\frac13\sqrt{m_s^2+2\tilde gH}$\\
$\Sigma^-$ & $dds$ & $\frac43m+\frac23m_s+\frac23\sqrt{m^2+2\tilde gH}+\frac13\sqrt{m_s^2+2\tilde gH}$\\
$\Xi^0$    & $ssu$ & $\frac23m+\frac43m_s+\frac13\sqrt{m^2+2\tilde gH}+\frac23\sqrt{m_s^2+2\tilde gH}$\\
$\Xi^-$    & $ssd$ & $\frac23m+\frac43m_s+\frac13\sqrt{m^2+2\tilde gH}+\frac23\sqrt{m_s^2+2\tilde gH}$\\
$\Lambda^0$& $uds$ & $\frac43m+\frac23m_s+\frac23\sqrt{m^2+2\tilde gH}+\frac13\sqrt{m_s^2+2\tilde gH}$\\
\hline\hline
\end{tabular}
\label{bm}
\end{table}

\begin{figure}[b]
\epsfig{file=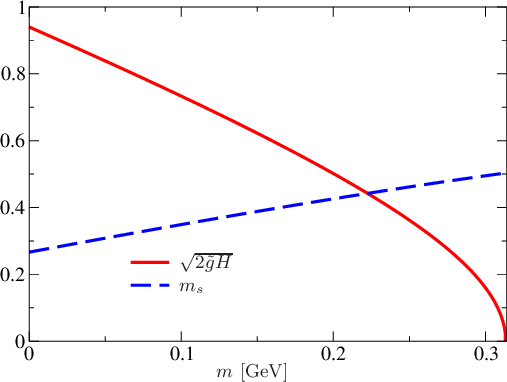,width=0.35\textwidth}
\caption{Red solid line: $\sqrt{2\tilde{g}H}$, blue dashed line: $m_s$. The numbers are given in units of  GeV.}
\label{msgH_m}
\end{figure}

Now, we estimate the value of $\tilde g H$ using the baryon masses as inputs.
Since we have three inputs, $940$ MeV for nucleons, $1130$ MeV for $m_\Lambda$ and $m_\Sigma$,
$1320$ MeV for $m_\Xi$, and three parameters to be determined, $m$, $m_s$, $\tilde g H$, we may be able to fix the value of $\tilde g H$.
However, all the baryon masses in the table are not independent from each other due to the Gell-Mann-Okubo mass relation~\cite{GOr}, and so we can at best find the relation among those three unknown parameters.
In the figure \ref{msgH_m}, we plot the dependence of $\sqrt{2{\tilde g}H}$ and $m_s$ on $m$. For example, if we take $m=300$ MeV, 
then we obtain $\sqrt{2\tilde gH}= 160$ MeV. As in Fig. 1, choosing a larger value of $m$ leads to a smaller value of $\sqrt{2\tilde gH}$.

To estimate the value of $gH$ and $m$ separately, we use the gluon condensate value as another input.
Using our background gluon field given in the section \ref{CCp}, we obtain the gluon condensate as a function of $gH$,
\begin{equation}
G_2=\left<\frac{\alpha_s}{\pi}G^{\mu\nu}_aG^a_{\mu\nu}\right>=\frac12\left(\frac{gH}{\pi}\right)^2.
\end{equation}
We use the three different values of the gluon condensate $G_2$ in Refs.~\cite{QCDsr, Horsley:2012ra, Bali:2014sja} compiled in Ref.~\cite{Gubler:2018ctz} to obtain the value of $gH$. 
We then use $M_N=2m+\sqrt{m^2+gH}$ and take $M_N=940\mathrm{\ MeV}$ to obtain the value of $m$. Results for these estimations of $gH$ and $m$ are shown in Table \ref{g2}. 

We put a cautionary remark here that  as is stated in \cite{Gubler:2018ctz}, the gluon condensate values presently are not much more than order of magnitude estimates with large uncertainties, and so do our estimated values of $gH$ and $m$ in Table \ref{g2}. 

The value of $m$ is negative when $G_2=0.077$ GeV$^4$, which can be explained as follows. 
By defining a new constituent quark field as $\tilde{\psi} = \gamma_5 \psi$, the Lagrangian (\ref{Lag-1}) is rewritten as
\ba
{\cal L}= \bar{\tilde{\psi}} i\not\!\!{D}\tilde{\psi} -\hat{m}\bar{\tilde{\psi}}\tilde{\psi}
-\frac{1}{2} F_{\mu\nu}F^{\mu\nu}+ \ldots\, ,
\label{Lag2}
\ea
where $\hat{m} = - m > 0$.
Then, the mass of the constituent quark described by the field $\tilde{\psi}$ is positive.

\begin{table}[!t]
\caption{Estimated values of $gH$ and $m$  based on the gluon condensate $G_2$ values.
For the negative value of the constituent quark mass, we refer to Eq. (\ref{Lag2}).}
\begin{tabular}{cc|c|c}
\hline\hline
$G_2$ [GeV$^4$] & Ref. & $\sqrt{gH}$ [MeV] & $m$ [MeV] \\ 
\hline
0.012 & \cite{QCDsr} &  697 &   117 \\
0.028 & \cite{Horsley:2012ra} &  862 &    39 \\
0.077 & \cite{Bali:2014sja} & 1110 & $-87$ \\
\hline\hline
\end{tabular}
\label{g2}
\end{table}

We also calculated the baryon decuplet masses and find that the masses of baryon octet and decuplet are the same: $M_N=M_\Delta$,
$M_\Sigma=M_{\Sigma^\star}$, $M_\Xi=M_{\Xi^\star}$.
Since we consider only the QCD vacuum effect, we can not explain the the mass difference between baryon octet and decuplet in our current study.
As well-known, the mass difference between baryon octet and decuplet (also between
pseudoscalar and vector mesons) is due to magnetic interactions, constituent quark spin-spin interactions.
 We note that the decuplet baryons constructed in our model include the effect of the gluon background field;
 Study of the effect to dense nuclear matter would be very interesting. (See e.g., Refs.~\cite{Takeda:2017mrm,Marczenko:2021uaj}.)

We finally remark here that apart from the contribution from the gluon background field to the baryon mass, there can be a contribution from the dynamical gluons that are supposed to interact weakly in the constituent quark model.
If we take the analogy of the Cornell potential,
$ V(r)=-\frac{a}{r} + b r $, the contribution from the background gluon field may correspond to the confinement part of the potential, while that from the dynamical gluons may correspond to the Coulomb part of the potential induced by one-gluon exchange.

\section{Summary and Discussion}\label{discussion}
To study the role of the QCD vacuum in nuclear and hadron physics, we have proposed to use the constituent quark model in the Copenhagen picture of the QCD vacuum which provides a center vortex scenario for confinement~\cite{Greensite:2011zz}.
Although we are using an effective model of QCD, our method could investigate the effects of QCD vacuum
 in terms of (constituent) quark and gluon degrees of freedom with less numerical computations. Especially, we can see the role of (background) gluon fields explicitly; in many cases the gluons are integrated out and encrypted in the low energy constants of an effective model or theory of QCD.

We have calculated the ground-state baryon masses and found that
the baryon mass is given by the sum of the effective constituent quark masses, $\tilde m$ and $\tilde m_s$, which depend
on the quantity that manifests the Copenhagen picture, $\sqrt{2{\tilde g}H}$,
 as shown in Eq.~(\ref{effective mass}).
We observed that the Gell-Mann-Okubo mass relation for the baryon octet is satisfied.
 In our current framework it turned out that the masses of baryon octet and decuplet (and also those of pseudoscalar and vector mesons)
 are the same.
 This is because we consider only the QCD vacuum effect, while the mass difference between baryon octet and decuplet is due to
  constituent quark spin-spin interactions via one-gluon exchange or Nambu-Goldstone-boson exchange.
We then estimated the values of $\sqrt{ gH}$ and $m_s$ as a function of
$u$ and $d$ constituent quark mass, $m$  by using the baryon masses as inputs. When $m= 300$ MeV, we obtained  $\sqrt{gH}=160$ MeV.

To be more realistic for the ground-state baryon properties such as
the mass difference between octet and decuplet baryons,
we need to include, for example,  one-gluon exchange  and/or meson exchanges in the Copenhagen vacuum.
A promising set up for this and also for other observables  might be the chiral quark model (ChQM) proposed in
\cite{Weinberg:1978kz, Manohar:1983md} with a suitable QCD vacuum such as the Copenhagen picture.

If we cast our picture into the ChQM~\cite{Manohar:1983md}, which is defined in between $\Lambda_{QCD}\sim 0.2$ GeV and $\Lambda_{\chi SB}\sim 1$ GeV, the interpretation about the smallness of the gauge coupling may change slightly.
In the ChQM, the large value of the gauge coupling near the scale $\Lambda_{\chi SB}$ leads to  chiral symmetry breaking
and so the gauge coupling can be small in a vacuum with chiral symmetry breaking.
In our case, we expect that the large value of the gauge coupling invokes
chiral symmetry breaking and develops a background gluon field or the Copenhagen picture  as we scale down to $\Lambda_{QCD}$. This way we may introduce QCD confinement to the ChQM.

As a future study, it will be interesting to investigate if one can see explicitly the effect of the QCD vacuum on physical quantities.
In our present work the effect of the QCD vacuum is  completely buried in the effective quark mass, and so
the baryon mass is simply the sum of the effective quark masses as in the Eq.(\ref{effective mass2}).
To see explicitly the effect of the QCD vacuum on physical quantities, we will consider 
physical quantities, other than the baryon masses, such as the pion decay constant. 
Another possibility might be that by including corrections to the baryon mass from dynamical gluons or pions, we may be able to
see explicitly the effect of the QCD vacuum on baryon masses.

It will be also informative to investigate if our approach, after a suitable extension, can provide a unified platform to deal with quark matter and nuclear matter on the same footing.

\newpage

\section*{Acknowledgments}
This work was supported partly by the
National Research Foundation of Korea (NRF) grant funded by the
Korea government (MSIT) (No.~2018R1A5A1025563), by the Institute for Basic Science (IBS-R031-D1),
and by JSPS KAKENHI Grant Numbers 20K03927, 23H05439.


\end{document}